\documentstyle[aps, manuscript]{revtex}
\tightenlines
\draft
\begin{document}
\title{de Sitter inflationary
expansion from a noncompact KK theory:
a nonperturbative quantum (scalar) field formalism}
\author{$^{1}$Mauricio Bellini\footnote{
E-mail address: mbellini@mdp.edu.ar}}
\address{$^1$Departamento de F\'{\i}sica, Facultad de
Ciencias Exactas y Naturales,
Universidad Nacional de Mar del Plata and
Consejo Nacional de Ciencia y Tecnolog\'{\i}a (CONICET),
Funes 3350, (7600) Mar del Plata, Argentina.}

\vskip .2cm
\maketitle
\begin{abstract}
We develop a nonperturbative quantum scalar field formalism
from a noncompact Kaluza-Klein (KK) theory using the
induced-matter theory of gravity during inflation.
We study the particular case
of a de Sitter expansion for the universe.
\end{abstract}
\vskip .2cm                             
\noindent
Pacs numbers: 04.20.Jb, 11.10.kk, 98.80.Cq \\
\vskip .1cm

\section{Introduction}

The two current versions of 5D gravity theory are membrane
theory\cite{au,aau,bu}
and induced-matter theory\cite{librowesson}.
In the former, gravity propagates freely
into the bulk, while the interactions of particle physics are confined
to a hypersurface (the brane).
The induced-matter theory in its simplest form is the basic
Kaluza-Klein (KK) theory in which the fifth dimension is not compactified
and the field equations of general relativity in 4D follow from the
fact that the 5D manifold is Ricci-flat; the large extra dimension is
thus responsible for the appearance of sources in 4D general relativity.
Hence, the 4D world of general relativity is embedded in a 5D Ricci-flat
manifold. An interesting result of the induced-matter theory is that if
$ds^2=g_{\mu\nu}(x) dx^{\mu} dx^{\nu}$ is the 4D metric of any
matter-free spacetime in
general relativity, the $dS^2 = \left(\psi/\psi_0\right)^2 ds^2 - d\psi^2$
is the metric of a 5D manifold that is
Ricci-flat\cite{MLW,MW}.

During the last two decades the inflationary paradigm has
become an almost universally accepted scenario to
explain the observed large scale flatness and homogeneity of the
universe\cite{lyth}.
In particular, stochastic inflation\cite{habib,mijic,copeland,liguori}
(or, in general, for
a semiclassical treatment for the inflaton field during inflation)
has been subject of great interest in the last years. However,
one of the problems with this approach
is that one must to
make a perturbative expansion of the scalar field potential
in terms of the quantum fluctuations of the inflaton field to
finally give a solution for a first order expansion in the equation
of motion for these fluctuations\cite{PRD}. This is a good
approximation because such that fluctuations are
small during inflation on cosmological scales. However, the
predictions of the inflationary theory could be significatively
improved by using a nonperturbative calculation for the
inflaton field $\varphi$. Of course, it is impossible to make from
a 4D quantum field formalism, but could be developed from a
scalar quantum (inflaton) field in a 5D vacuum state with
a purely kinetic density Lagrangian.
The aim of this work consists to develop a nonperturbative
scalar quantum field theory from a 5D apparent vacuum state
defined as a purely kinetic 5D density Lagrangian of a scalar
field minimally coupled to gravity in a 5D Ricci-flat canonical
metric\cite{MB}.
To make it, we consider the 
5D canonical metric\cite{PLB}
\begin{equation}\label{6}
dS^2 = \psi^2 dN^2 - \psi^2 e^{2N} dr^2 - d\psi^2,
\end{equation}
where $dr^2=dx^2+dy^2+dz^2$. 
Here, the coordinates ($N$,$\vec r$)
are dimensionless and the fifth coordinate
$\psi $ has spatial unities.
We shall assume in what follows that the extra dimension is
spacelike and that the universe is 3D spatially flat, isotropic and
homogeneous.
The metric (\ref{6}) describes a
flat 5D manifold in apparent vacuum ($G_{AB}=0$).
We consider a diagonal metric because we are dealing only with
gravitational effects, which are the important ones during
inflation.
To describe neutral matter in a 5D geometrical vacuum
(\ref{6}) we can consider the Lagrangian
\begin{equation}\label{1}
^{(5)}{\rm L}(\varphi,\varphi_{,A}) =
-\sqrt{\left|\frac{^{(5)}
g}{^{(5)}g_0}\right|} \  ^{(5)}{\cal L}(\varphi,\varphi_{,A}),
\end{equation}
where $|^{(5)}g|=\psi^8 e^{6N}$
is the absolute value of the determinant for the 5D metric tensor with
components $g_{AB}$ ($A,B$ take the values $0,1,2,3,4$) and
$|^{(5)}g_0|=\psi^8_0 e^{6N_0}$
is a constant of dimensionalization determined
by $|^{(5)}g|$ evaluated at $\psi=\psi_0$ and $N=N_0$.
In this work we shall consider $N_0=0$, so that
$^{(5)}g_0=\psi^8_0$.
Here, the index ``$0$'' denotes the values at the end of inflation.
Furthermore, we shall consider an action
\begin{displaymath}
I = - {\Large\int} d^4x d\psi \sqrt{\left|\frac{^{(5)}g}{^{(5)}g_0}\right|}
\left[\frac{^{(5)} R}{16\pi G} + {\cal L}(\varphi,\varphi_{,A})\right],
\end{displaymath}
where $\varphi$ is a scalar field minimally coupled to gravity
and $G$ is the gravitational constant.
Furthermore, $^{(5)} R$ is the 5D Ricci scalar, which of course, is cero for
the 5D flat metric (\ref{6}).

Since the 5D metric (\ref{6}) describes a manifold in apparent
vacuum, the density Lagrangian
${\cal L}$ in (\ref{1}) must to be
\begin{equation}\label{1'}
^{(5)}{\cal L}(\varphi,\varphi_{,A}) = 
\frac{1}{2} g^{AB} \varphi_{,A} \varphi_{,B},
\end{equation}
which represents a free scalar field. In other words, we define the vacuum
as a purely kinetic 5D-lagrangian on a globally 5D-flat metric [in our
case, the metric (\ref{6})].
To describe
the metric in physical coordinates we can make the
following transformations:
\begin{equation}
t = \psi_0 N, \qquad R= \psi_0 r, \qquad \psi= \psi,
\end{equation}
such that we obtain the 5D metric
\begin{equation}\label{m1}
dS^2 = \left(\frac{\psi}{\psi_0}\right)^2 \left[dt^2
- e^{2t/\psi_0} dR^2\right]- d\psi^2,
\end{equation}
where $t$ is the cosmic time and $R^2=X^2+Y^2+Z^2$.
This metric is the Ponce de Leon one\cite{...}, and describes a
3D spatially flat, isotropic and homogeneous extended (to 5D)
FRW metric in a de Sitter expansion\cite{librowesson}.

\section{Quantum field theory in a 5D apparent vacuum}

Taking into account
the metric (\ref{6}) and the Lagrangian (\ref{1}), we
obtain the equation of motion for $\varphi$
\begin{equation}\label{df}
\left(2\psi \frac{\partial\psi}{\partial N}+ 3 \psi^2 \right)
\frac{\partial\varphi}{\partial N}
+\psi^2 \frac{\partial^2\varphi}{\partial N^2}
- \psi^2 e^{-2N} \nabla^2_r\varphi
-4\psi^3 \frac{\partial\varphi}{\partial\psi} - 3\psi^4 \frac{\partial N}{
\partial\psi} \frac{\partial\varphi}{\partial\psi} - \psi^4 \frac{\partial^2
\varphi}{\partial\psi^2} =0,
\end{equation}
where ${\partial N \over \partial\psi}$ is zero because the coordinates
$(N,\vec{r},\psi)$ are independents. Hence, we obtain
\begin{equation}
\stackrel{\star\star}{\varphi} + 3 \stackrel{\star}{\varphi} -
e^{-2N} \nabla^2_r \varphi - \left[4\psi \frac{\partial\varphi}{\partial
\psi} + \psi^2 \frac{\partial^2\varphi}{\partial\psi^2}\right] =0,
\end{equation}
where the overstar denotes the derivative with respect to $N$ and
$\varphi \equiv \varphi(N,\vec R,\psi)$. To simplify its structure
we can make the transformation $\varphi = \chi e^{-3N/2} \left({\psi_0\over
\psi}\right)^2$, so that we obtain the 5D generalized Klein-Gordon
like equation for the redefined field $\chi(N,\vec r, \psi)$:
\begin{equation}
\stackrel{\star\star}{\chi} - \left[ e^{-2N} \nabla^2_r + \left(
\psi^2 \frac{\partial^2}{\partial\psi^2} + \frac{1}{4} \right)\right]
\chi =0.
\end{equation}
The field $\chi$ can be written in terms of a Fourier expansion
\begin{equation}\label{9}
\chi(N,\vec r, \vec{\psi}) = \frac{1}{(2\pi)^{3/2}} {\Large\int} d^3k_r
{\Large\int} dk_{\psi} \left[ a_{k_r k_{\psi}} e^{i (\vec{k_r}.\vec r+
\vec{k_{\psi}}.\vec{\psi})} \xi_{k_r k_{\psi}}\left(N,\psi\right) +
a^{\dagger}_{k_r k_{\psi}} e^{-i (\vec{k_r}.\vec r+
\vec{k_{\psi}}.\vec{\psi})} \xi^*_{k_r k_{\psi}}\left(N,\psi\right)\right],
\end{equation}
where the asterisk denotes the complex conjugate and
($a^{\dagger}_{k_r k_{\psi}}$,$a_{k_r k_{\psi}}$) are the creation and
annihilation operators such that
\begin{eqnarray}
&& \left[a_{k_r k_{\psi}}, a^{\dagger}_{k'_r k'_{\psi}}\right] =
\delta^{(3)}\left(\vec k_r - \vec k'_r\right) \  \delta\left(\vec{k_{\psi}} -
\vec{k'_{\psi}}\right). \\
&& \left[a^{\dagger}_{k_r k_{\psi}}, a^{\dagger}_{k'_r k'_{\psi}}\right] =
\left[a_{k_r k_{\psi}}, a_{k'_r k'_{\psi}}\right] =0.
\end{eqnarray}
Furthermore, the commutation relation between $\chi$ and
$\stackrel{\star}{\chi}$ is
\begin{equation}\label{co}
\left[\chi(N,\vec r, \psi), \stackrel{\star}{\chi}(N,\vec r',\psi')\right]=
i \delta^{(3)}\left(\vec r - \vec r'\right) \  \delta\left(
\vec{\psi} - \vec{\psi'}\right).
\end{equation}
In order to the commutation equation (\ref{co}) holds
the following
renormalization condition must to be fulfilled:
\begin{equation}   \label{re1}
\xi_{k_r k_{\psi}} \left(\stackrel{\star}{\xi}_{k_r k_{\psi}}\right)^* -
\left(\xi_{k_r k_{\psi}}\right)^* \stackrel{\star}{\xi}_{k_r k_{\psi}} =i.
\end{equation}
Hence, the equation for the modes $\xi_{k_r k_{\psi}}(N,\psi)$
that complies with the condition (\ref{re1}) in a 4D de Sitter
expansion will be
\begin{equation}\label{re}
\stackrel{\star\star}\xi_{k_r k_{\psi}} +
\left[ e^{-2N} k^2_r -\left(\frac{1}{4} - \psi^2 k^2_{\psi}\right)\right]
\xi_{k_r k_{\psi}} =0.
\end{equation}
The general solution for this equation is
\begin{equation}
\xi_{k_r k_{\psi}}\left(N,\psi\right) = G_1(\psi) \
{\cal H}^{(1)}_{\nu}\left[
k_r e^{-N}\right] + G_2(\psi) \  {\cal H}^{(2)}_{\nu}\left[k_r e^{-N}\right],
\end{equation}
where ${\cal H}^{(1,2)}_{\nu}[x] = {\cal J}_{\nu}[x] \pm i {\cal Y}_{\nu}[x]$
are the
Hankel functions, ${\cal J}_{\nu}[x]$ and
${\cal Y}_{\nu}[x]$ are the first and sencond kind
Bessel functions, and $\nu = {\sqrt{1-4k^2_{\psi} \psi^2} \over 2}$.
Furthermore, the functions $G_1(\psi)$ and $G_2(\psi)$ are
arbitrary functions constrained by the renormalization condition (\ref{re1})
\begin{equation}
\left[G_1(\psi) - G_2(\psi)\right]\left[G_1(\psi) + G_2(\psi)\right] =
\frac{\pi}{4}.
\end{equation}
In this paper we shall choose the generalized Bunch-Davis vacuum:
$G_1(\psi)=0$ and $G_2(\psi) =i {\sqrt{\pi} \over 2}$.
The squared $\varphi$ fluctuations are given by
\begin{equation}
\left<\varphi^2\right> = \frac{1}{(2\pi)^3} {\Large\int} d^3k_r {\Large\int}
dk_{\psi} \  \xi_{k_rk_{\psi}} \xi^*_{k_rk_{\psi}}.
\end{equation}
A $k_r$-scale invariant power spectrum results from $\nu = 3/2$, for which
\begin{equation}
i k_{\psi} = -\frac{\sqrt{2}}{\psi}.
\end{equation}

\section{4D de Sitter expansion}

We can take a foliation $\psi=\psi_0$ in the metric
(\ref{m1}), such that the effective 4D metric
results
\begin{equation}\label{m2}
dS^2 \rightarrow ds^2 = dt^2 - e^{2t/\psi_0} dR^2,
\end{equation}
which describes 4D globally isotropic and homogeneous expansion of
a 3D spatially flat, isotropic and homogeneous universe that expands
with a Hubble parameter $H=1/\psi_0$ (in our case a constant)
and a 4D scalar curvature
$^{(4)}{\cal R} = 6(\dot H + 2 H^2)$. Note that in this particular
case the Hubble parameter is constant so that $\dot H = 0$.

The 4D energy 
density $\rho$ and the pressure ${\rm p}$ are\cite{PLB}
\begin{eqnarray}
&& 8 \pi G \left<\rho\right> = 3 H^2,\\
&& 8\pi G \left<{\rm p}\right> = -3H^2,
\end{eqnarray}
where $G=M^{-2}_p$ is the gravitational constant and
$M_p=1.2 \  10^{19} \  GeV$ is the Planckian mass.
Furthermore, the universe describes a vacuum
equation of state: ${\rm p} = - \rho$,
such that
\begin{equation}
\left<\rho\right> = \left<\frac{\dot\varphi^2}{2} + \frac{a^2_0}{2a^2}
\left(\vec\nabla \varphi\right)^2 + V(\varphi)\right>,
\end{equation}
where the brackets denote the 4D expectation vacuum
and the cosmological constant $\Lambda $ gives the vacuum energy density
$\left<\rho\right> = {\Lambda \over 8\pi G}$. Thus, $\Lambda $ is
related with the fifth coordinate by means of $\Lambda = 3/\psi^2_0$\cite{...}.
Furthermore, the 4D Lagrangian is given by
\begin{equation}
^{(4)}{\cal L}(\varphi,\varphi_{,\mu}) =
-\sqrt{\left|\frac{^{(4)}g}{^{(4)}g_0}\right|} \left[
\frac{1}{2} g^{\mu\nu} \varphi_{,\mu}\varphi_{,\nu} + V(\varphi)\right],
\end{equation}
where the effective potential for the 4D FRW metric\cite{MB}, is
\begin{equation}\label{pot}
V(\varphi) = -\left.\frac{1}{2} g^{\psi\psi} \varphi_{,\psi}\varphi_{,\psi}
\right|_{\psi=\psi_0} =
\frac{1}{2} \left.
\left(\frac{\partial\varphi}{\partial\psi}\right)^2\right|_{\psi=\psi_0}
\end{equation}
In our case this potential takes the form
\begin{equation}\label{pot2}
V(\varphi) = \left(\frac{2}{\psi_0^2} - \frac{k^2_{\psi_0}}{2} -
\frac{2 i k_{\psi_0}}{\psi_0} \right) \varphi^2(t,\vec R,\psi_0),
\end{equation}
where $k_{\psi_0}$ is the wavenumber for $\psi=\psi_0$.
Furthermore the effective 4D motion equation for $\varphi$ is
\begin{equation}
\ddot\varphi + \frac{3}{\psi_0} \dot\varphi -
e^{-2t/\psi_0} \nabla^2_R\varphi - \left.\left[
4\frac{\psi}{\psi^2_0} \frac{\partial\varphi}{\partial\psi} +
\frac{\psi^2}{\psi^2_0} \frac{\partial^2\varphi}{\partial\psi^2}\right]
\right|_{\psi=\psi_0} =0,
\end{equation}
which means that the effective derivative (with respect to $\varphi$)
for the potential, is
\begin{equation}
\left.V'(\varphi)\right|_{\psi=\psi_0} =  \alpha\varphi(\vec R,t,\psi_0),
\end{equation}
with
\begin{equation}
\alpha = \frac{2}{\psi^2_0}+k^2_{\psi_0}.
\end{equation}
Now we can make the following transformation:
\begin{equation}
\varphi(\vec R, t) = e^{-\frac{3t}{2\psi_0}} \chi (\vec R, t).
\end{equation}
Note that now $\varphi\equiv \varphi(\vec R=\psi_0 r,t=\psi_0 N, \psi=\psi_0)=
e^{-3t/(2\psi_0)} \chi(\vec R, t)$, where [see eq. (\ref{9})]
$\chi(\vec R, t) = \chi(\vec R=\psi_0 r, t=\psi_0 N, \psi=\psi_0)$:
\begin{equation}
\chi(\vec R, t) = \frac{1}{(2\pi)^{3/2}}
{\Large\int} d^3 k_R {\Large\int} dk_{\psi} \left[
a_{k_R k_{\psi}} e^{i(\vec{k_R}.\vec{ R} + \vec{k_{\psi}}.\vec{\psi})}
\xi_{k_R k_{\psi}}(t,\psi) + c.c.
\right] \  \delta\left( k_{\psi} - k_{\psi_0}\right).
\end{equation}
Hence, we obtain the following 4D Klein-Gordon equation
for $\chi $
\begin{equation}   \label{mod}
\ddot\chi - \left[e^{-\frac{2t}{\psi_0}} \nabla^2_R +\frac{9}{4\psi^2_0}
-\alpha\right]\chi =0.
\end{equation}
The equation of motion for the time dependent
modes $\xi_{k_Rk_{\psi_0}}(t)$ will
be
\begin{equation}\label{mot}
\ddot\xi_{k_Rk_{\psi_0}} + \left[ k^2_R e^{-\frac{2t}{\psi_0}} 
-\left(\frac{9}{4\psi^2_0}
-\alpha\right)\right]\xi_{k_R k_{\psi_0}} =0,
\end{equation}
where the effective squared mass $\mu^2$ of the modes $\xi_{k_R k_{\psi_0}}$
is given by
\begin{equation}                    \label{mu}
\mu^2 = \frac{9}{4\psi^2_0} - \alpha.
\end{equation}
Here, $\alpha $ describes the self-interaction squared mass
of the redefined
inflaton field $\chi$ due to the expansion of the universe, and the
term ${9 \over 4\psi^2_0}$ represents its bare squared mass due to
the expansion. Note that both terms in (\ref{mu})
has a geometrical origin because they
are induced by the fifth coordinate.

The general solution of eq. (\ref{mot}) is
\begin{equation}\label{sol}
\xi_{k_Rk_{\psi_0}} =
G_1(\psi_0) \  {\cal H}^{(1)}_{\frac{3-n_s}{2}}\left[k_R \psi_0 e^{-t/\psi_0}\right]
+ G_2(\psi_0) \  {\cal H}^{(2)}_{\frac{3-n_s}{2}}\left[k_R \psi_0 e^{-t/\psi_0}\right].
\end{equation}
Here, ${(3-n_s) \over 2}=\sqrt{{9\over 4} -(2+k^2_{\psi_0}\psi^2_0)}$
and $n_s$ is the spectral index of the $\left<\varphi^2\right>$-spectrum
on super Hubble (SH) scales when
it is considered the Bunch-Davis vacuum: $G_1(\psi_0)=0$,
$G_2(\psi_0)=i\sqrt{\pi}/2$
\begin{equation}
\left.\left<\varphi^2\right> \right|^{SH} =
{\Large\int}^{\epsilon k_0}_{k_*} \frac{dk}{k} {\cal P}(t,k) \simeq
A(t) {\Large\int}^{\epsilon k_0}_{k_*} \frac{dk}{k} \  k^{n_s},
\end{equation}
where $A(t)$ is a time dependent function,
$k_0(t) = {e^{t/\psi_0}\over \psi_0} \sqrt{{9\over 4} -\alpha\psi^2_0}$,
$\epsilon \ll 1$, ${\cal P}(t,k)$ is the power of the spectrum
and $k_*$ is the absolute value for the wave vector
related to the physical wavelength at the moment the horizon entry.
This maximal physical scale is supported by causal arguments.
Note that this result corresponds to an effective self-interaction
squared mass $\alpha $
for the inflaton field
\begin{equation}
\alpha = 2 H^2 +k^2_{\psi_0}; \qquad {\rm for \  \psi_0=1/H},
\end{equation}
when the standard semiclassical approach in a de Sitter expansion
is considered\cite{PRD}.
For $n_s \ll 1$, it can be approximated to
$n_s\simeq {2(2+k^2_{\psi_0}\psi^2_0) \over 3}$, where
\begin{equation}\label{eq}
i k_{\psi_0} \simeq -\frac{\sqrt{2-3 n_s/2}}{\psi_0}
= -\sqrt{2-3 n_s/2} \  H.
\end{equation}
It is well known that the universe has a power spectrum which
is very close to a scale invariant one
on cosmological scales\cite{prl}, so that the equation (\ref{eq}) is
a good approximation.
In such that case we obtain the following valued expressions
for $V(\varphi)$ and
$V'(\varphi)$:
\begin{eqnarray}
&& \left.V(\varphi)\right|_{\left(\psi=\psi_0, n_s=0\right)}
= \frac{(4+2\sqrt{2})}{\psi^2_0}
\varphi^2(\vec R, t, \psi_0),\\ \label{vv}
&& \left.V'(\varphi)\right|_{\left(\psi=\psi_0, n_s=0\right)}
=0. \label{v'}
\end{eqnarray}
where the equation (\ref{v'}) corresponds to $\alpha =0$ in
(\ref{mod}).
Therefore, $\alpha$ becomes zero for a super Hubble scale invariant
power spectrum of $\left<\varphi^2\right>$.
Note that this result disagrees with the result
obtained using a semiclassical 4D treatment for the inflaton
field in a de Sitter expansion\cite{PRD}.
Furthermore, we can see from eq. (\ref{vv}) that the effective
4D parameter of mass for $\varphi$ 
\begin{equation}\label{mm}
m^2_{eff} = \frac{2(4+2\sqrt{2})}{\psi^2_0},
\end{equation}
is nonzero. It describes the inflaton $\varphi$-squared mass geometrically
induced by the fifth coordinate $\psi$ on the hypersurface
$\psi=\psi_0$, when the self-interaction is ausent in a de Sitter expansion:
$\alpha =0$.
It is easy to see that $m^2_{eff} > \left.\mu^2\right|_{\alpha=0}$.
However, both masses has a different origin, because $\mu^2$ is the squared
mass related to each $\xi_{k_R k_{\psi_0}}$ mode of $\chi$ and
$m^2_{eff}$ is the effective squared mass of the
nonperturbative field $\varphi(\vec R,t,\psi_0)$,
with back-reaction effects included.

\subsection{Energy density fluctuations}

Once the time dependent modes $\xi_{k_Rk_{\psi_0}}$ are known, we
can to obtain the effective 4D expectation value for the energy density
\begin{eqnarray}
\left<\rho\right>_{\psi=\psi_0}
&=&
\frac{e^{-3t/\psi_0}}{(2\pi)^3}
{\Large\int}^{k_p}_{k_*}
d^3k_R \left\{\left[\frac{\left(26+(3-n_s)^2\right)}{8\psi^2_0}
- \frac{
2 i k_{\psi_0}}{\psi_0} + \frac{e^{-2t/\psi_0} k^2_R}{2}\right]
\xi_{k_R k_{\psi_0}} \xi^*_{k_R k_{\psi_0}}\right. \nonumber \\
&+& \left.\frac{1}{2} \dot\xi_{k_R k_{\psi_0}}
\dot\xi^*_{k_R k_{\psi_0}} - \frac{3}{4\psi_0} \left(\xi_{k_R k_{\psi_0}}
\dot\xi^*_{k_R k_{\psi_0}}
+ \dot\xi_{k_R k_{\psi_0}}\xi^*_{k_R k_{\psi_0}}\right)\right\},
\end{eqnarray}
where $k_p = G^{-1/2}$ is the absolute value of the Planckian
wave vector and $k_*$ is the inverse
of the Hubble's radius at the moment ($t=t_*$) the
horizon entry.

On the other hand, if we define $\left<\rho\right>^{(0)}_{\psi=\psi_0}$ the
$k_R$-zero mode 4D expectation value for the energy density 
\begin{eqnarray}
\left<\rho\right>^{(0)}_{\psi=\psi_0}
&=&
\frac{e^{-3t/\psi_0}}{(2\pi)^3}
{\Large\int}^{k_p}_{k_*}
d^3k_R \left\{\left[\frac{\left(26+(3-n_s)^2\right)}{8\psi^2_0}
- \frac{2 i k_{\psi_0}}{\psi_0} \right]
\xi_{0 k_{\psi_0}} \xi^*_{0 k_{\psi_0}}\right. \nonumber \\
&+& \left.\frac{1}{2}\dot\xi_{0 k_{\psi_0}}
\dot\xi^*_{0 k_{\psi_0}} - \frac{3}{4\psi_0} \left(\xi_{0k_{\psi_0}}
\dot\xi^*_{0 k_{\psi_0}}
+ \dot\xi_{0 k_{\psi_0}}\xi^*_{0 k_{\psi_0}}\right)\right\},
\end{eqnarray}
on the effective 4D spatially isotropic and homogeneous (de Sitter)
FRW metric (\ref{m2}), hence the energy density fluctuations
(on arbitrary scales) will be
\begin{equation}\label{den}
\delta\rho/\rho = {<\rho>_{\psi=\psi_0} - <\rho>^{(0)}_{\psi=\psi_0} \over
<\rho>_{\psi=\psi_0}}.
\end{equation}

\subsection{Super Hubble energy density fluctuations}

It is well known that the universe is nearly scale invariant
on cosmological scales\cite{prl}: $|n_s| \ll 1$.
We shall consider this range for the spectrum on the range
$k_* < k_R < \epsilon k_0(t)$.
In order to compute the SH density energy fluctuations we can
remember the small-argument limit for the second kind Hankel function:
${\cal H}^{(2)}_{\nu}[x] \simeq -{i\over\pi} \Gamma(\nu) (x/2)^{-\nu}$,
for $\nu >0$, which in our case is valid to describe the super Hubble
(cosmological) $\varphi$-field time dependent modes
$\xi_{k_r k_{\psi_0}}$ in eq. (\ref{sol}). With this approach, we obtain
\begin{eqnarray}
\xi_{k_R k_{\psi_0}} \xi^*_{k_R k_{\psi_0}} & \simeq &
\frac{2^{1-n_s}}{\pi} \Gamma^2\left(\frac{3-n_s}{2}\right) \  \left[
k_R \psi_0 e^{-t\psi_0}\right]^{-(3-n_s)}, \\
\dot\xi_{k_R k_{\psi_0}} \dot\xi^*_{k_R k_{\psi_0}} & \simeq &
\left(\frac{3-n_s}{2}\right)^2 \frac{\Gamma^2\left(\frac{3-n_s}{2}\right)}{
2^{1+n_s} \pi \psi^2_0} \left[
k_R \psi_0 e^{-t/\psi_0}\right]^{-(1-n_s)}, \\
\frac{\xi_{k_R k_{\psi_0}} \dot\xi^*_{k_R k_{\psi_0}}
+ \dot\xi_{k_R k_{\psi_0}}\xi^*_{k_R k_{\psi_0}}}{2} 
& \simeq &
\left(\frac{3-n_s}{2}\right) \frac{\Gamma^2\left(\frac{3-n_s}{2}\right)}{
2^{n_s} \pi \psi_0} \left[
k_R \psi_0 e^{-t/\psi_0}\right]^{-(2-n_s)},
\end{eqnarray}
and for the zero-modes, we obtain
\begin{eqnarray}
\xi_{0 k_{\psi_0}} \xi^*_{0 k_{\psi_0}} & \simeq &
\frac{2}{\pi} \Gamma^2\left(\frac{3}{2}\right) \left[
k_R \psi_0 e^{-t/\psi_0}\right]^{-3}, \\
\dot\xi_{0 k_{\psi_0}} \dot\xi^*_{0 k_{\psi_0}} & \simeq &
\frac{9}{8\pi} \frac{\Gamma^2\left(\frac{3}{2}\right)}{\psi^2_0}\left[
k_R \psi_0 e^{-t/\psi_0}\right]^{-1}, \\
\frac{\xi_{0 k_{\psi_0}} \dot\xi^*_{0 k_{\psi_0}}
+ \dot\xi_{0 k_{\psi_0}}\xi^*_{0 k_{\psi_0}}}{2} 
& \simeq &
\frac{3}{2\pi } \frac{\Gamma^2\left(\frac{3}{2}\right)}{\psi_0}\left[
k_R \psi_0 e^{-t/\psi_0}\right]^{-2}.
\end{eqnarray}
Hence, on cosmological scales (i.e., for $k^2 \ll k^2_0$), the
late times
energy density fluctuations renormalized at $n_s=0$ will be
approximately (i.e., at first order in $n_s$)
\begin{equation}\label{fin}
\left.\frac{\delta\rho}{\rho}\right|^{SH}_{t=t_*} \simeq
\left. n_s  {\rm ln}\left[k_R\right]\right|^{\epsilon k_0(t_*)}_{k_*},
\end{equation}
where $t_*$ is the time when the horizon entry and
we have made use of the expression (\ref{eq}).
Note that eq. (\ref{fin}) describes the departure of $\delta\rho/\rho$
on super Hubble scales with respect to scale invariant
density fluctuations ($n_s=0$).
Furthermore, the expression (\ref{fin}) is consistent for
\begin{equation}
n_s \  {\rm ln}\left.\left[k_R\right]
\right|^{\epsilon k_0(t_*)}_{k_*} < 1.
\end{equation}
Hence, if $k_* e^{-H_* t_*} = H_*$ is the Hubble parameter when
the horizon entry and we consider
${\epsilon \sqrt{{9\over 4} - \alpha\psi^2_0}\over H_* \psi_0}
e^{\left(\psi^{-1}_0 - H_*\right)t_*} > e^{60}$, one obtains the
following inequality\footnote{In the calculations we are considering
$\psi^{-1}_0 = H$, in consistence with the 4D FRW metric (\ref{m2})}:
\begin{equation}
60 < {\rm ln}\left.\left[k_R\right]\right|^{\epsilon k_0(t_*)}_{k_*}
< \frac{1}{n_s},
\end{equation}
which is fulfilled for
\begin{equation}
n_s < 1/60.
\end{equation}
This is an important result which is in good agreement with
observation\cite{PDB}.
\section{Final Comments}

We have developed a 5D nonperturbative quantum scalar field formalism
and studied the particular case of a 4D de Sitter inflationary expansion
in the framework of the induced-matter theory developed by Wesson and
co-workers.
The results are very interesting, because we note some differences
with respect to whose obtained by means of the semiclassical expansion
in a de Sitter expansion for the universe. 1) First all, the
effective scalar 4D potential appears to be quadratic in $\varphi$ (see eq.
(\ref{pot2}),
meanwhile in the standard semiclassical treatment
is a constant $V_0$ (see for example
\cite{PRD}). 2) The spectral index $n_s < 1/60$ depends on the
self-interaction squared
mass $\alpha$ (which should be very small if the slow-roll conditions
hold), but not on the mass of the inflaton field $m_{eff}$ (as in
the standard semiclassical approach to inflation).
Notice this result suggests that $k_{\psi_0} \simeq
i \sqrt{2}/\psi_0 = i \sqrt{2} H$.
3) Furthermore, the Hubble parameter in a de Sitter expansion
(as in the semiclassical approach to inflation) is a constant, but in our
case, its value being given by the inverse of the fifth coordinate
$H=1/\psi_0$ for a foliation $\psi=\psi_0$
of the 5D Ricci-flat metric (\ref{m1}).

Finally, the nonperturbative treatment for $\varphi$ here
developed can be extended to other (more realistic)
inflationary models
with time dependent Hubble parameters and, more generally, to cosmological
models dinamically governed by scalar fields.\\

\vskip .3cm
\noindent
M.B. Acknowledges CONICET, AGENCIA and UNMdP (Argentina) for financial
support.\\

\end{document}